# EXPERIMENTS AND NUMERICAL RESULTS ON NONLINEAR VIBRATIONS OF AN IMPACTING HERTZIAN CONTACT.

# PART 2: RANDOM EXCITATION.

J. P<small>ERRET</small>-L<small>IAUDET AND</small> E. R<small>IGAUD</small>

*Laboratoire de Tribologie et Dynamique des Systèmes UMR 5513,*

*36 avenue Guy de Collongue, 69134 Ecully cedex, France.*

Short running title: RANDOMLY EXCITED HERTZIAN CONTACT

22 pages

5 tables

13 figures

Final version

**Summary**

Non linear dynamic behaviour of a normally excited preloaded Hertzian contact (including possible contact losses) is investigated using an experimental test rig. It consists on a double sphere plane contact loaded by the weight of a rigid moving mass. Contact vibrations are generated by a external Gaussian white noise and exhibit vibroimpact responses when the input level is sufficiently high. Spectral contents and statistics of the stationary transmitted normal force are analysed. A single-degree-of-freedom non linear oscillator including loss of contact and Hertzian non linearities is built for modelling the experimental system. Theoretical responses are obtained by using the stationary Fokker-Planck equation and also Monte Carlo simulations. When contact loss occurrence is very occasional, numerical results shown a very good agreement with experimental ones. When vibroimpacts occur, results remain in reasonable agreement with experimental ones, that justify the modelling and the numerical methods described in this paper.

The contact loss non linearity appears to be rather strong compared to the Hertzian non linearity. It actually induces a large broadening of the spectral contents of the response. This result is of great importance in noise generation for a lot of systems such as mechanisms using contacts to transform motions and forces (gears, ball-bearings, cam systems, to name a few). It is also of great importance for tribologists preoccupied to prevent surface dammage.

# 1. Introduction

Hertzian contacts exist in many mechanical systems such as mechanisms and machines (gears, cam systems, rolling element bearings, …). Under operating conditions, these contacts are often excited by dynamic normal forces superimposed on a mean static load. These excitation forces which are deterministic or random, can result from external sources, such as applied load fluctuations, or from internal ones, such as roughness-induced vibrations. Under excessive excitation, contacts can exhibit undesirable vibroimpact responses, allowed by clearances introduced through manufacturing tolerances. Resulting dynamic behaviour is characterised by intermittent loss of contact and shocks leading to excessive wear, surface damage and excessive noise.

In a companion paper (Part I: Harmonic excitation) [12], the dynamic behaviour of a fundamental preloaded Hertzian contact subjected to harmonic normal forces was studied. To this end, an improved experimental test rig permitted us to investigate the primary resonance in detail, including vibroimpact responses. Theoretical results were also presented to conclude on the main characteristics of the primary resonance.

In the present second part of this work, the previous analysis is extended to the case of vibroimpact response of a preloaded and non-sliding dry Hertzian contact under Gaussian white random normal excitation. Comparisons between experimental and theoretical results permit us to conclude on some characteristics of the random dynamic responses, including vibroimpact behaviours.

As the literature shows, there exist a few number of papers in this area. These include references [1-7]. They generally concern random normal vibrations for sliding contacts related to the internal random roughness surface induced excitation.

Experimental results can be found for example in references [2-5;7]. In these studies, excitation random force applied to the contact is never exactly known because measurements are always performed during sliding conditions. It is estimated on the basis of assumptions on the spatial spectrum of the surface roughness input. Decrease of the frequency domain spectrum shape with a $\omega^{-4}$ law is generally retained. These works principally focus on the interaction between normal and tangential forces, friction force vibrations and friction coefficient behaviour under dynamic conditions. Further, vibroimpact behaviours are very partially analysed.

In a theoretical point of view, Nayak [1] presents a detailed analysis of a Hertzian contact excited by a broadband random normal force. But, intermittent loss of contact is partially taken into account. This problem is examined by Hess et al. [5] who consider also small values of probability of contact loss. The used theoretical method is based on the Fokker-Planck equation, introducing the restoring elastic Hertzian force by a third order Taylor expansion. The procedure is refined by Pärssinen in a recent brief paper by introducing the real form of the theoretical restoring elastic Hertzian force [6].

## 2. Test rig and experimental procedure

The experimental studied system is similar to the one presented in part I of this work [12]. Recall that it consists on a 25.4 mm diameter steel ball preloaded between two horizontal steel flat surfaces. The first one is rigidly fixed to a vertically moving cylinder, the second one is rigidly fixed to a heavy rigid frame. The double sphere-plane dry contact is loaded by a static normal load $F_s = mg = 110$ N which corresponds to the weight of the moving cylinder.

Using a suspended vibration exciter, random normal force is applied to the moving cylinder and superimposed on the static load. For this end, a signal generator and a power amplifier are used to generate the white noise signal. This signal is filtered above 1 kHz which is up to 4 times the experimental linearised contact frequency ($f_0 = 233.4$ Hz).

A piezoelectric force transducer is mounted between the vibration exciter and the moving cylinder to measure the excitation force. Normal force N(t) transmitted to the base through the contact is measured by a piezoelectric force transducer mounted between the lower plane and the rigid frame. Considering the experimental system and the transducer stiffness (8000 N/µm), the force measurement bandwidth is sufficiently wide (0-7 kHz).

The input force and the dynamic response are displayed on a storage oscilloscope. Experimental average one-sided RMS magnitude spectra are measured with a real time spectrum analyser using a sampling rate of 4096 samples over the frequency bandwidth (0-1 kHz) with a frequency resolution always less than 0.25 Hz. Average spectra were obtained with a number of spectrum up to 275. Dynamic responses are digitised using an A/D converter and stored for statistical post-treatments. These responses are sampled for a 20 s duration with sampling rate of 10000 samples per second.

## 3. The theoretical dynamic model

### 3.1 Equation of motion

On the basis of identical assumptions that ones introduced in the first part of this work, the experimental system is modelled by a randomly excited single-degree-of-freedom non-linear dynamic system shown in Figure 1 and described by the following motion equation:

$$m\ddot{z} + c\dot{z} + k[z\,H(z)]^{3/2} = F_S(1 + \sqrt{h}\,W(t)) \qquad (1)$$

In this equation, z is the normal displacement of the rigid mass m measured such as $z < 0$ corresponds to loss of contact. Assuming viscous law, c is a damping coefficient, k is a constant given by the Hertzian theory, H is the Heaviside step function, W(t) is a stationary zero-mean Gaussian white noise and h controls level of the random normal force. From equation (1), the theoretical static contact compression $z_S$ and the linearised contact natural circular frequency $\Omega$ are given by:

$$z_s = \left(\frac{F_s}{k}\right)^{2/3} \qquad \Omega^2 = \left(\frac{3k}{2m}\right) z_s^{1/2} \qquad (2,3)$$

Recall that data have been previously found from the experimental characteristics as:

$$z_s = 7 \, \mu m, \qquad f_0 = \frac{\Omega}{2\pi} = 232 \, Hz \qquad (4,5)$$

Now, by letting:

$$q = \frac{3}{2}\left(\frac{z - z_s}{z_s}\right), \qquad \tau = \Omega t, \qquad \varpi = \omega/\Omega \qquad (6,7,8)$$

dimensionless equation of motion is achieved in the same way as in part I:

$$\ddot{q} + 2\zeta\dot{q} + [(1 + \frac{2}{3}q)H(1 + \frac{2}{3}q)]^{3/2} = 1 + \sqrt{h}w(\tau) \qquad (9)$$

In this equation, overdot indicates differentiation with respect to the dimensionless time $\tau$, $\zeta$ is an equivalent viscous damping ratio, parameter h controls the input level and $w(\tau)$ is chosen as a stationary zero-mean Gaussian white noise with a unit power spectral density, $S_{ww}(\varpi) = 1$. So, considering the dimensionless excitation force $f(\tau) = \sqrt{h}w(\tau)$, the power spectral density and autocorrelation function are given by:

$$S_{ff}(\varpi) = h \qquad (10)$$

$$R_{ff}(\tau) = 2\pi h \delta(\tau) \qquad (11)$$

where $\delta(\tau)$ is the Dirac function. With this choice, one should notice that the power spectral density of $W(t)$ is equal to $1/\Omega$. Hence, power spectral density of the excitation force $F(t) = F_S\sqrt{h}W(t)$ is equal to:

$$S_{FF}(\omega) = S_0 = F_S^2 h/\Omega \qquad (12)$$

From the experimental one sided spectral density $G_{FF}(f) = G_0$, we have:

$$G_0 = 4\pi S_0 \qquad (13)$$

$$h = \frac{f_0 G_0}{2F_S^2} \qquad (14)$$

Finally, it should be noted from equation (9) that loss of contact corresponds to the inequality:

$$q < -3/2 \tag{15}$$

**3.2 Numerical methods**

To investigate theoretical dynamic responses, we have used several numerical methods as follows:

- a classical numerical time integration explicit scheme, i.e. the central difference scheme, for achieving dynamic time histories of responses,

- the stationary Fokker-Planck equation applied to system (6) under Gaussian white-noise excitation for describing the statistics of the dynamic responses,

- classical statistical tools to describe random vibrations,

- Monte Carlo simulations to estimate power spectral densities of responses of the randomly excited system.

When necessary, these methods are described in the following.

**4. Experimental results**

**4.1. Dynamics without loss of contact**

Figure 2 displays the experimental average one-sided RMS magnitude spectra of the transmitted normal force for various random input levels. These levels are chosen in such a way that no loss of contact occurs during the measure. For these cases, $G_0 = 1 \; 10^{-4}$, $G_0 = 4.5 \; 10^{-3}$ and $G_0 = 30 \; 10^{-3}$ N$^2$/Hz with corresponding values of h = $1 \; 10^{-6}$, $4.5 \; 10^{-5}$ and $3 \; 10^{-4}$ respectively. For the lowest input level, the spectrum exhibits a single resonant peak close to the linearised contact frequency (233.4 Hz). This result is in very good agreement with the predicted one (232 Hz). In this case, linear behaviour can be assumed and an equivalent viscous damping ratio can be estimated from the half power frequency bandwidth method. Experimental curve leads to $\zeta$ less than 0.5 %. This result is coherent with the previous value found in the part I of this work [12].

When the input level increases, dynamic behaviour becomes weakly non-linear. Actually, a second peak close to the second harmonic of the linearised contact frequency arises. For the highest input level shown in Figure 2, the second peak value reaches 8 % of the main peak value. Furthermore, increasing the input level, we observe a weak broadening of the two peaks. We also observe weak decrease of the two peak frequencies.

Experimental normal force probability density functions corresponding to the three preceding input levels are presented in Figure 3. The associated mean, standard deviation, and skewness values are given in Table 1. The normal force shows a nearly zero probability of intermittent contact loss (which corresponds to N<-1). Further, increasing the input level, the Gaussian-like shape of the probability density function moves to a more and more asymmetrical shape. This is confirmed by the evolution of the skewness value given in Table 1. However, the deviation from a Gaussian process is weak. Considering the mean value of the normal force, it remains also very close to the applied static load. Further, the standard deviation reaches 40 % of the static normal force.

Figure 4 displays an example of the time history of the normal force for an input level such that intermittent contact loss does not occur during acquisition data. In agreement with the spectral responses content, this trace is typical of a narrow-frequency-band random noise response. A slight asymmetrical response can be observed revealing the asymmetrical law of the restoring Hertzian elastic force around the static load.

### 4.2. Dynamics with intermittent loss of contact

When the random input level increases, intermittent contact loss may occur as the normal force can reach the static load. Figure 5 shows the experimental average one-sided RMS magnitude spectra for increasing random input levels, chosen in such a way that intermittent loss of contact occurs during the measure. For these cases, $G_0 = 6\ 10^{-2}$, $G_0 = 1.7\ 10^{-1}$ and $G_0 = 2.5\ 10^{-1}$ N$^2$/Hz with corresponding values of $h = 6\ 10^{-4}$, $1.7\ 10^{-3}$ and $2.5\ 10^{-3}$ respectively. The broadening of the resonant peaks is clearly observed. It is known that it is an essential property of spectrum shapes with large non-linearity and low damping [9]. One can also observe the rising of a broad third peak. Furthermore, resonant peaks shift to lower frequencies, particularly the second and the third ones. This can be explained by the more and more asymmetrical shape of the time trace of the transmitted force induced by the flight response of the cylinder. Hence, as expected, loss of contact non-linearity is stronger than the Hertzian contact one.

Figure 6 displays the corresponding experimental normal force probability density functions. The associated mean value, standard deviation and skewness values are given in Table 2. Probability density functions are strongly asymmetrical and so deviate hardly from a Gaussian process. This is confirmed by Skewness values given in Table 2. Asymmetrical behaviour corresponds to the appearance of a peak at the zero transmitted normal force in the probability density functions (N=-1). In particular, the longer flight time of the cylinder, the higher the peak is. Calculation of the peak area allows estimation of the total loss of contact duration (this has been also estimated by treating time histories with coherent results). Results are reported in Table 3. For the highest input level, intermittent contact loss occurs during approximately 15 % of the overall time.

Figure 7 displays an example of the time history of the normal force for an input level such that intermittent contact loss occurs. From time traces, we measure the total number of contact losses during the acquisition time (20 s). We estimate also the mean period of the cylinder flight. Results are presented in Table 3. For the highest input level, this mean period is approximately equal to 1.5 ms and the number of impacts becomes large (around 2200 during a observation time of 20 s).

## 5. Theoretical results

### 5.1. The stationary Fokker-Planck equation

Assuming stationary Gaussian white noise excitation, statistics of the stationary response can be obtained using the Fokker-Planck equation [1,5,6,10]. To this end, consider the following second order non linear differential equation of motion, in the general form:

$$\ddot{q} + 2\zeta\dot{q} + G(q) = f(\tau) \qquad (16)$$

where $f(\tau)$ is a zero mean stationary Gaussian white noise excitation with intercorrelation function $R_{ff}(\tau) = 2\pi h\, \delta(\tau)$, i.e. a power spectral density $S_{ff}(\varpi) = h$. $G(q)$ represents the non linear restoring force including the non-linearity related to contact loss. The forward Fokker-Planck equation which governs the transitional probability density function $p(q,\dot{q},t|q_0,\dot{q}_0)$ of system (16) is obtained as follows:

$$\frac{\partial p}{\partial t} + \dot{q}\frac{\partial p}{\partial q} = \frac{\partial}{\partial \dot{q}}[2\zeta\dot{q}p + G(q)p] + \pi h \frac{\partial^2 p}{\partial \dot{q}^2} \qquad (17)$$

Considering the stationary case and after some rearrangements, the stationary joint probability density function $p_s(q,\dot{q})$ satisfies:

$$\frac{\partial}{\partial \dot{q}}[G(q)p_s + \frac{\pi h}{2\zeta}\frac{\partial p_s}{\partial q}] + (2\zeta\frac{\partial}{\partial \dot{q}} - \frac{\partial}{\partial q})[\dot{q}p_s + \frac{\pi h}{2\zeta}\frac{\partial p_s}{\partial \dot{q}}] = 0 \qquad (18)$$

Hence, a solution is achieved by requiring:

$$[G(q)p + \frac{\pi h}{2\zeta}\frac{\partial p}{\partial q}] = 0 \qquad (19)$$

and

$$[\dot{q}p + \frac{\pi h}{2\zeta}\frac{\partial p}{\partial \dot{q}}] = 0 \qquad (20)$$

From (19) and (20), one easily obtains a solution for the stationary joint probability density function $p_s(q,\dot{q})$ as:

$$p_s(q,\dot{q}) = A\exp\left\{\frac{-2\zeta}{\pi h}\frac{\dot{q}^2}{2}\right\}\exp\left\{\frac{-2\zeta}{\pi h}\int_0^q G(s)ds\right\} \qquad (21)$$

where A is a constant which normalises the density function. From (21), marginal densities for the displacement and the velocity appear statistically independents and a closed form for the displacement probability density function $p_q(q)$ is easily achieved as follows:

$$p_q(q) = C\exp\left\{\frac{-2\zeta}{\pi h}\int_0^q G(s)ds\right\} \qquad (22)$$

or from expression of G(q):

$$\begin{cases} p_q(q) = C\exp\left\{-\frac{2\zeta}{\pi h}\left[\frac{3}{5}(1+\frac{2}{3}q)^{5/2}-q-\frac{3}{5}\right]\right\} & \text{if } q > -\frac{3}{2} \\ p_q(q) = C\exp\left\{-\frac{2\zeta}{\pi h}\left[-q-\frac{3}{5}\right]\right\} & \text{if } q \le -\frac{3}{2} \end{cases} \qquad (23)$$

where C is a constant which normalises the marginal density function.

As one can see in equation (21), marginal density function for the velocity is a Gaussian process.

$$p_{\dot{q}}(\dot{q}) = B\exp\left\{\frac{-2\zeta}{\pi h}\frac{\dot{q}^2}{2}\right\} \qquad (24)$$

where B is a constant which normalises the marginal density function.

By using numerical integration methods, statistical moments of the displacement $q(\tau)$ are computed from the probability density function (23).

$$E[q^n(\tau)] = \int_{-\infty}^{+\infty} q^n p_q(q)dq \qquad (25)$$

Since the relation between the displacement $q(\tau)$ and the elastic restoring force $N(\tau)$ is known, the probability density function of the elastic restoring force $p_N(N)$ is derived in a classical manner [11].

Consider the relation N = G(q):

$$\begin{cases} N = (1+\frac{2}{3}q)^{3/2} - 1 & \text{if } q > -\frac{3}{2} \\ N = -1 & \text{if } q \leq -\frac{3}{2} \end{cases} \quad (26)$$

If $q_i$, $i = 1..r$, are all real roots of (26), one obtains the probability density function of the elastic restoring force as:

$$p_N(N) = \sum_{i=1}^{r} \frac{p_q(q_i)}{|N'(q_i)|} \quad (27)$$

where N'(q) is the derivative of N with respect to q.

Finally:

$$\begin{cases} p_N(N) = (1+N)^{-1/3} p_q\left(\frac{3}{2}(1+N)^{2/3} - \frac{3}{2}\right) & \text{if } N > -1 \\ p_N(-1) = \delta(-1) \int_{-\infty}^{-3/2} p_q(s) ds \\ p_N(N) = 0 & \text{if } N < -1 \end{cases} \quad (28)$$

where $\delta$ is the Dirac delta function. Notice that $p_N$ is zero for $N < -1$, and contains an impulse at $N = -1$ of area equal to the probability of loss of contact. Notice also that:

$$\lim_{\substack{N \to -1 \\ N > -1}} p_N(N) = +\infty \quad (29)$$

Statistical moments are given as follows:

$$E[N^n(\tau)] = \int_{-\infty}^{+\infty} N^n p_N(N) dN \quad (30)$$

from which mean value, standard deviation and skewness values can be derived.

### 5.2. Monte Carlo simulations

We have performed Monte Carlo simulations to estimate response spectral densities of the randomly excited system. For this end, we have used an explicit numerical time integration scheme (central difference method) for solving the motion equation (9) and for achieving dynamic time histories of the normal force. Spectra were obtained via a Fast Fourier Transform procedure with a number of samples equal to 2048. Average spectra were obtained with a number of spectrum up to 400. To simulate the Gaussian white noise external force, we have considered a sufficiently wide band limited pseudo-random signal given by:

$$w(\tau) = W_M \sum_{k=1}^{k=M} \cos(\varpi_k \tau + \phi_k) \qquad (31)$$

Frequencies $\varpi_k$ are independent and uniformly distributed in $]0,\varpi_{max}]$ and $W_M$ is a coefficient which take into account the frequency resolution.

$$W_M = \sqrt{\Delta f \cdot h} \qquad (32)$$

Choosing phases $\phi_k$ as follows:

$$\phi_k = 2\pi\sqrt{-2r_1}\cos(2\pi r_2) \qquad (33)$$

where $r_1$ and $r_2$ are two random numbers uniformly distributed over [0,1], the signal is normally distributed.

Also, Monte Carlo simulations have been used to obtain statistics of the stationary responses treating the theoretical time histories. Statistical moments have been estimated over $10^5$ times the period of the linearised system with up to 250 samples by period.

### 5.3. Theoretical results

### 5.3.1 Statistics of the stationary response

The probability densities of the elastic restoring force are shown in Figure 8 (h = 3 $10^{-6}$, 8 $10^{-5}$, 5 $10^{-4}$) and Figure 9 (h = 1.2 $10^{-3}$, 2 $10^{-3}$, 3.2 $10^{-3}$) for increasing input levels and for a damping ratio equal to 0.5%. The two figures correspond respectively to input levels chosen in such a way that no loss and loss of contact occurs. Both results derived from the Fokker-Planck equation and the Monte Carlo simulations are shown. As we can see, a good agreement between the two methods is observed. Small differences observed on Figure 9 can result from numerical errors associated to the estimation of the Dirac value at N=−1. By comparing Figure 8 and Figure 3, we conclude on a good agreement between experimental and numerical results. Actually, the introduced input levels leading to the same probability density function shapes are not exactly the same but appear to be of the same order. Moreover, it should be stated that the pertinent variable is the ratio between damping ratio and input level (see equation 21). Further, damping law as well as damping value are not exactly known particularly when the amplitude response grows. So, in our opinion, adjusting damping ratio gives no more satisfaction than adjusting input level. However, one can assume that the experimental damping ratio of 0.5 % is overestimated. Actually, by taking into account the average ratio between the two series of input levels, we conclude that a damping ratio equal to 0.3 % is more convenient. By comparing Figure 9 and Figure 6, we conclude on a satisfactory agreement between experimental and numerical results when intermittent contact losses occur. Of course, the theoretical infinite value and Dirac function at $N = -1$ cannot be observed in the experimental case. Further, ball motion between the two planes associated to the second mode and clearly experimentally observed (see Figure 7) is a source of discrepancy between theoretical and experimental probability density functions. Numerical statistics obtained from equation (25) are given in Table 4. Statistical results obtained from Monte Carlo simulations are reported in Table 5. Concerning standard deviations, satisfactory agreement is obtained between measured and computed results (see Tables 1 and 2). In accordance with the preceding remarks, we have found that this agreement becomes very good if we scale the input level by damping ratio with a value equal to 0.3 % . So, we can conclude that the numerical tools used in this study are suitable for describing standard deviations, even if the probability density function are not exactly the same when contact losses occur. Concerning mean values, discrepancy between experimental results and numerical ones is found. However, it should be pointed out that the numerical mean values remain very close to the static applied load which is coherent with experimental data. Further, samplings of both experimental time traces and those obtained from Monte Carlo simulations are perhaps not sufficient to ensure precise estimate of the mean value. Concerning skewness values, good agreement between experimental and Monte Carlo simulations results is obtained, but discrepancies appear with results obtained from the Fokker-Planck equation. We don't have precise explanation, but we can assume that sampling procedure leads numerical errors for estimating skewness values. Finally, probabilities of contact loss given in Table 4 have been computed from equation (23). Again, very good agreement with experimental results is observed (compare the total loss of contact duration given in Table 3).

### 5.3.2 Response spectra

Figures 10 and 11 display the numerical average one-sided RMS magnitude spectra for increasing input levels (h = $3\ 10^{-6}$, $8\ 10^{-5}$, $5\ 10^{-4}$, $1.2\ 10^{-3}$, $2\ 10^{-3}$, $3.2\ 10^{-3}$). Comparisons with Figures 2 and 5 reveals a very good agreement between experimental and numerical results when intermittent contact losses do not occur. The agreement remains satisfactorily when contact losses occur, even if experimental level of the second and third peaks are higher than those obtained from numerical simulations.

Since the experimental test rig does not permit higher input levels, it can be interesting to know the effect of increasing input level on the transmitted force response through numerical simulations. Figure 12 displays the average one-sided RMS magnitude spectra of the elastic restoring force for input levels higher than the preceding ones. The result is a large broadening of the spectral density response, in such a way that preceding peaks completely disappear.

### 5.3.3 Effect of the Hertzian contact law on response spectra

Since linear restoring force law is introduced in a lot of modelling of practical systems, it is of interest to compare results between modellings which include the Hertzian contact law or not. For this end, we have consider a simplified model for describing the elastic restoring force as follows:

$$N = (1+q).H(1+q) - 1 \qquad (34)$$

Figure 13 displays the numerical average one-sided RMS magnitude spectra for three input levels (h = 8 $10^{-5}$, 2 $10^{-3}$, 1.4 $10^{-2}$) showing comparisons between results obtained from the two models. As one can see, identical results are obtained around the primary peak. In contrast, the simplified model is not suitable to describe the dynamic behaviour in the range of higher frequencies. Depending on the objective, approximation (34) can be favourably introduced under cover of saving of numerical time consuming.

## 6. Conclusion

An experimental test rig consisting on a dry double sphere-plane Hertzian contact is modelled as a nonlinear single-degree-of-freedom system. Nonlinear normal force response of this randomly excited system is analysed through experimental and theoretical results.

For very low input force amplitude, almost linear behaviour is observed, and the experimental linearised contact frequency is deduced with a very good agreement with the predicted one. Equivalent viscous damping ratio is less than 0.5 %. Increasing the input amplitude reveals the second harmonic spectral peak. Probability density functions of the response show a little deviation from a Gaussian process resulting from the Hertzian non-linearity as long as intermittent contact loss is very occasional.

As the input level increases, the probability of intermittent loss of contact strongly increases. The probability density functions of the response become largely asymmetrical. Furthermore, we observe the rising of the third harmonic spectral peak and the broadening of spectral peaks. This last behaviour is known as an essential property of the power spectral density of systems with large non-linearity and low damping. It is clearly verified in our experimental results.

For all these behaviours, numerical and experimental results agree. We conclude that the associated theoretical model is sufficiently accurate, despite the fact that damping is modelled in a very simple way. So, compared to vibrations under harmonic excitation, precise knowledge of the damping law during vibroimpact response is less determining. Also, stationary Fokker-Planck equation and Monte Carlo simulations are suitable methods for describing the dynamic behaviour of the impacting Hertzian contact under normal random excitation.

In future, further numerical and experimental works are planned to take into account the effect of a lubricating film and the effect of sliding surfaces.

## 7. Acknowledgement

## 9. Nomenclature

m rigid moving mass

c damping coefficient

k constant obtained from Hertzian theory

$F_s$ static load

F(t) excitation normal force

W(t) Gaussian white noise process

h parameter controlling input level

$z(t)$ normal displacement

$z_s$ static contact compression

$\Omega$ linearised natural circular frequency

$f_0$ linearised natural frequency

$\zeta$ damping ratio

$\tau$ dimensionless time

$q(\tau)$ dimensionless normal displacement

$f(\tau)$ dimensionless excitation normal force

$w(\tau)$ dimensionless Gaussian white noise process

$\omega$ circular frequency

$\varpi$ dimensionless circular frequency

$N(\tau)$ Hertzian elastic restoring force

$S_{xx}(\omega)$ power spectral density of x

$G_{xx}(\omega)$ one-sided power spectral density of x

$R_{xx}(\tau)$ intercorrelation function of x

$p_x(x)$ probability density function of x

$E[x]$, $\bar{x}$ mean value of x

$\sigma$, standard deviation

$\gamma$, skewness value

Table 1. Mean value E[N], standard deviation σ, and skewness γ of the measured dimensionless transmitted force.

Table 2. Mean value E[N], standard deviation σ and skewness γ of the measured dimensionless transmitted force.

Table 3. Loss of contact duration.

Table 4. Mean value E[N], standard deviation σ and skewness γ of the computed dimensionless transmitted force by using the Fokker-Planck equation.

Table 5. Mean value E[N], standard deviation σ and skewness γ of the computed dimensionless transmitted force by using Monte Carlo simulations.

Figure 1. The studied randomly excited single-degree-of-freedom oscillator.

Figure 2. Experimental one sided RMS spectra of the transmitted normal force for $h \approx 1\ 10^{-6}$, $4.5\ 10^{-5}$ and $3\ 10^{-4}$ (respectively a, b, c).

Figure 3. Probability density functions of the measured transmitted normal force for $h \approx 1\ 10^{-6}$, $4.5\ 10^{-5}$ and $3\ 10^{-4}$ (respectively a, b, c).

Figure 4. Time traces of the transmitted normal force $h \approx 1\ 10^{-6}$, $4.5\ 10^{-5}$ and $3\ 10^{-4}$ (respectively a, b, c).

Figure 5. Experimental one sided RMS spectra of the transmitted normal force for $h \approx 6\ 10^{-4}$, $1.7\ 10^{-3}$ and $2.5\ 10^{-3}$ (respectively a, b, c).

Figure 6. Probability density functions of the measured transmitted normal force for $h \approx 6\ 10^{-4}$, $1.7\ 10^{-3}$ and $2.5\ 10^{-3}$ (respectively a, b, c).

Figure 7. Time traces of the transmitted normal force for $h \approx 6\ 10^{-4}$, $1.7\ 10^{-3}$ and $2.5\ 10^{-3}$ (respectively a, b, c).

Figure 8. Probability density functions of the elastic restoring force for $h = 3\ 10^{-6}$, $8\ 10^{-5}$ and $5\ 10^{-4}$ (respectively a, b, c). Results obtained from the stationary Fokker-Planck equation (——) and from Monte Carlo simulations ( O ).

Figure 9. Probability density functions of the elastic restoring force for $h = 1.2\ 10^{-3}$, $2\ 10^{-3}$, $3.2\ 10^{-3}$ (respectively a, b, c). Results obtained from the stationary Fokker-Planck equation (——) and from Monte Carlo simulations ( O ).

Figure 10. Numerical one sided RMS spectra of the elastic restoring force for $h = 3\ 10^{-6}$, $8\ 10^{-5}$ and $5\ 10^{-4}$ (respectively a, b, c).

Figure 11. Numerical one sided RMS spectra of the elastic restoring force for $h = 1.2\ 10^{-3}$, $2\ 10^{-3}$ and $3.2\ 10^{-3}$ (respectively a, b, c).

Figure 12. Numerical one sided RMS spectra of the elastic restoring force for $h = 7\ 10^{-3}$, $1.4\ 10^{-2}$ and $3\ 10^{-1}$ (respectively a, b, c).

Figure 13. Numerical one sided RMS spectra of the elastic restoring force obtained with an Hertzian elastic contact law (thin line) and a linear one (thick line). $h = 8\ 10^{-5}$, $2\ 10^{-3}$ and $1.4\ 10^{-2}$ (respectively a, b, c).

| h | E[N] | σ | γ |
|---|---|---|---|
| $1.0\ 10^{-6}$ | $-9.8\ 10^{-5}$ | 0.03 | 0.011 |
| $4.5\ 10^{-5}$ | $7.0\ 10^{-6}$ | 0.15 | 0.089 |
| $3.0\ 10^{-4}$ | $4.6\ 10^{-5}$ | 0.38 | 0.260 |

Table 1. Mean value E[N], standard deviation σ, and skewness γ of the measured dimensionless transmitted force.

J. PERRET-LIAUDET AND E. RIGAUD

| h | E[N] | σ | γ |
|---|---|---|---|
| $6.0 \times 10^{-4}$ | $-2.1 \times 10^{-4}$ | 0.70 | 0.406 |
| $1.7 \times 10^{-3}$ | $-3.4 \times 10^{-4}$ | 0.78 | 0.464 |
| $2.5 \times 10^{-3}$ | $-2.4 \times 10^{-4}$ | 1.01 | 0.758 |

Table 2. Mean value E[N], standard deviation σ and skewness γ of the measured dimensionless transmitted force.

| h | Total loss of contact duration | Nunber of loss of contact during 20 s | Mean period of cylinder fly |
|---|---|---|---|
| $6.0 \; 10^{-4}$ | 3 % | 693 | 0.9 ms |
| $1.7 \; 10^{-3}$ | 6 % | 1050 | 1.1 ms |
| $2.5 \; 10^{-3}$ | 15 % | 2227 | 1.5 ms |

Table 3. Loss of contact duration.

| h | E[N] | σ | γ | p(q<–1.5) |
|---|---|---|---|---|
| $3\ 10^{-6}$ | $-1.57\ 10^{-3}$ | 0.03 | - 0.010 | ≈ 0 % |
| $8\ 10^{-5}$ | $-4.23\ 10^{-3}$ | 0.16 | - 0.054 | ≈ 0 % |
| $5\ 10^{-4}$ | $-2.81\ 10^{-2}$ | 0.40 | - 0.157 | ≈ 0 % |
| $1.2\ 10^{-3}$ | $-1.33\ 10^{-3}$ | 0.60 | 0.502 | 2 % |
| $2.0\ 10^{-3}$ | $-1.66\ 10^{-3}$ | 0.75 | 0.710 | 7 % |
| $3.2\ 10^{-3}$ | $-2.14\ 10^{-3}$ | 0.91 | 0.951 | 15 % |

Table 4. Mean value E[N], standard deviation σ and skewness γ of the computed dimensionless transmitted force by using the Fokker-Planck equation.

| h | E[N] | σ | γ |
|---|---|---|---|
| $3\,10^{-6}$ | $-3.0\,10^{-4}$ | 0.03 | 0.018 |
| $8\,10^{-5}$ | $-1.0\,10^{-3}$ | 0.15 | 0.097 |
| $5\,10^{-4}$ | $-2.0\,10^{-3}$ | 0.38 | 0.260 |
| $1.2\,10^{-3}$ | $-4.1\,10^{-3}$ | 0.57 | 0.490 |
| $2.0\,10^{-3}$ | $-4.2\,10^{-3}$ | 0.72 | 0.666 |
| $3.2\,10^{-3}$ | $-4.4\,10^{-3}$ | 0.88 | 0.917 |

Table 5. Mean value E[N], standard deviation σ and skewness γ of the computed dimensionless transmitted force by using Monte Carlo simulations.

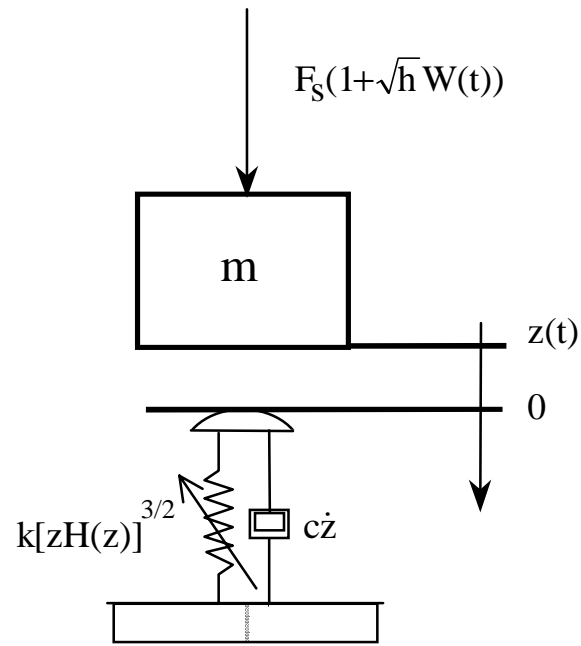

Figure 1. The studied randomly excited single-degree-of-freedom oscillator.

J. PERRET-LIAUDET AND E. RIGAUD

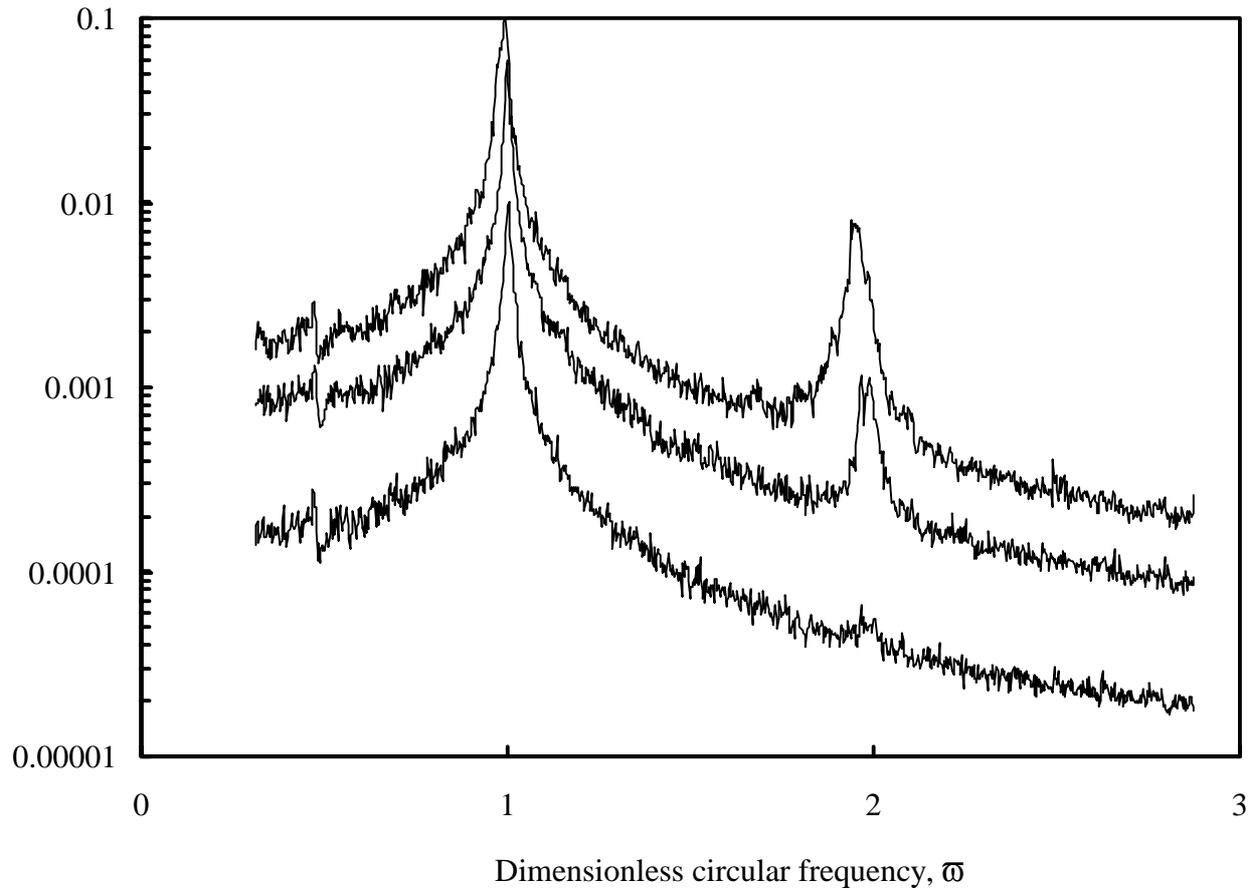

Figure 2. Experimental one sided RMS spectra of the transmitted normal force for h ≈ $1\ 10^{-6}$, $4.5\ 10^{-5}$ and $3\ 10^{-4}$ (respectively a, b, c).

J. PERRET-LIAUDET AND E. RIGAUD

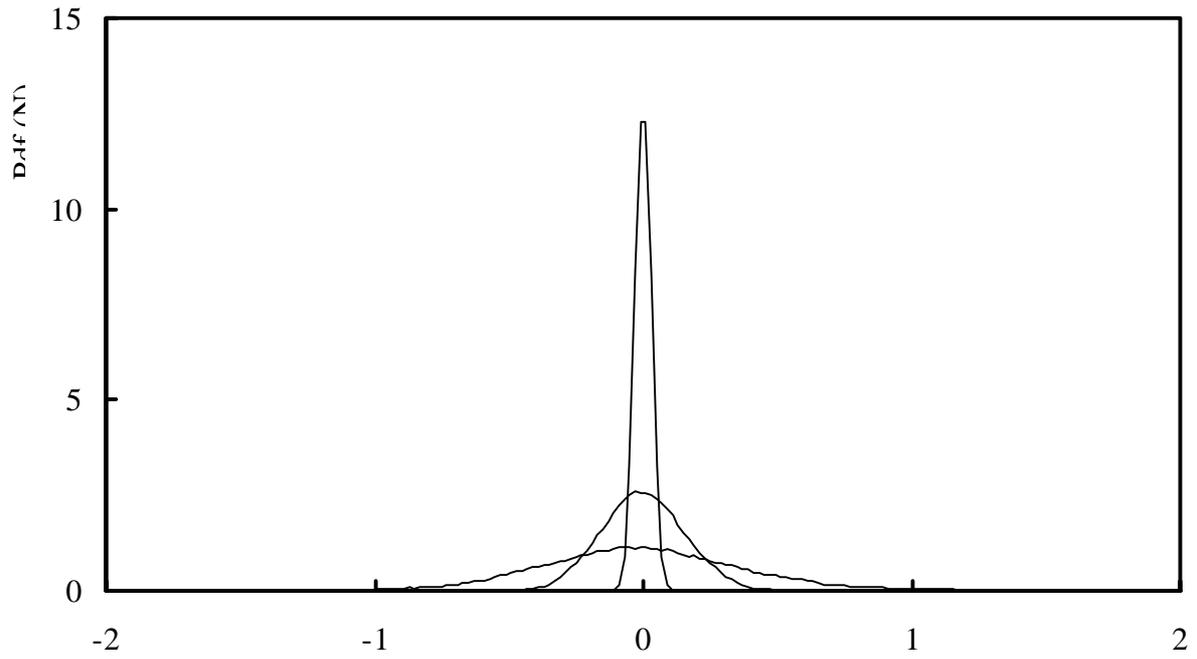

Figure 3. Probability density functions of the measured transmitted normal force for h ≈ $1\ 10^{-6}$, $4.5\ 10^{-5}$ and $3\ 10^{-4}$ (respectively a, b, c).

J. PERRET-LIAUDET AND E. RIGAUD

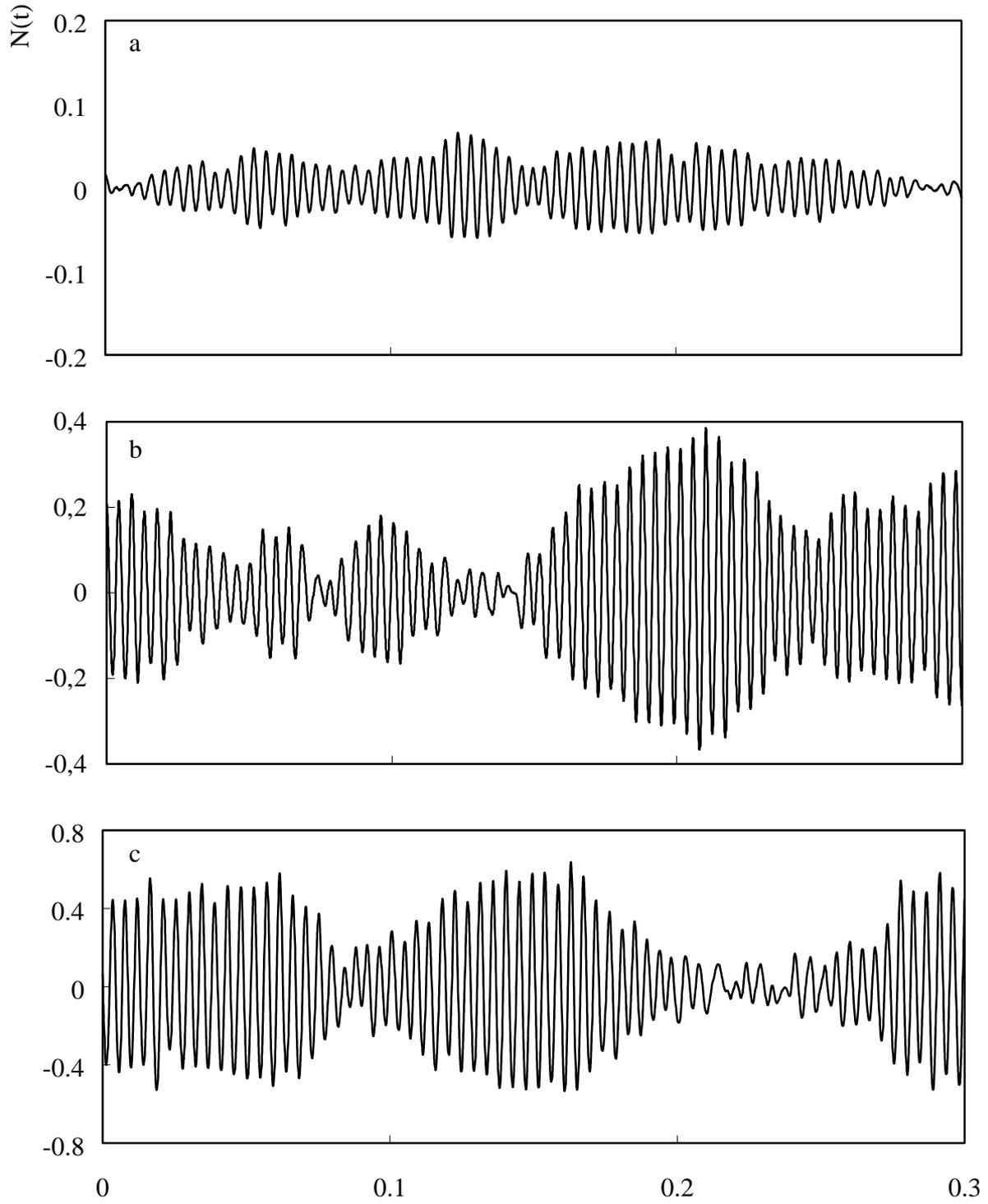

Figure 4. Time traces of the transmitted normal force h ≈ $1\ 10^{-6}$, $4.5\ 10^{-5}$ and $3\ 10^{-4}$ (respectively a, b, c).

J. Perret-Liaudet and E. Rigaud

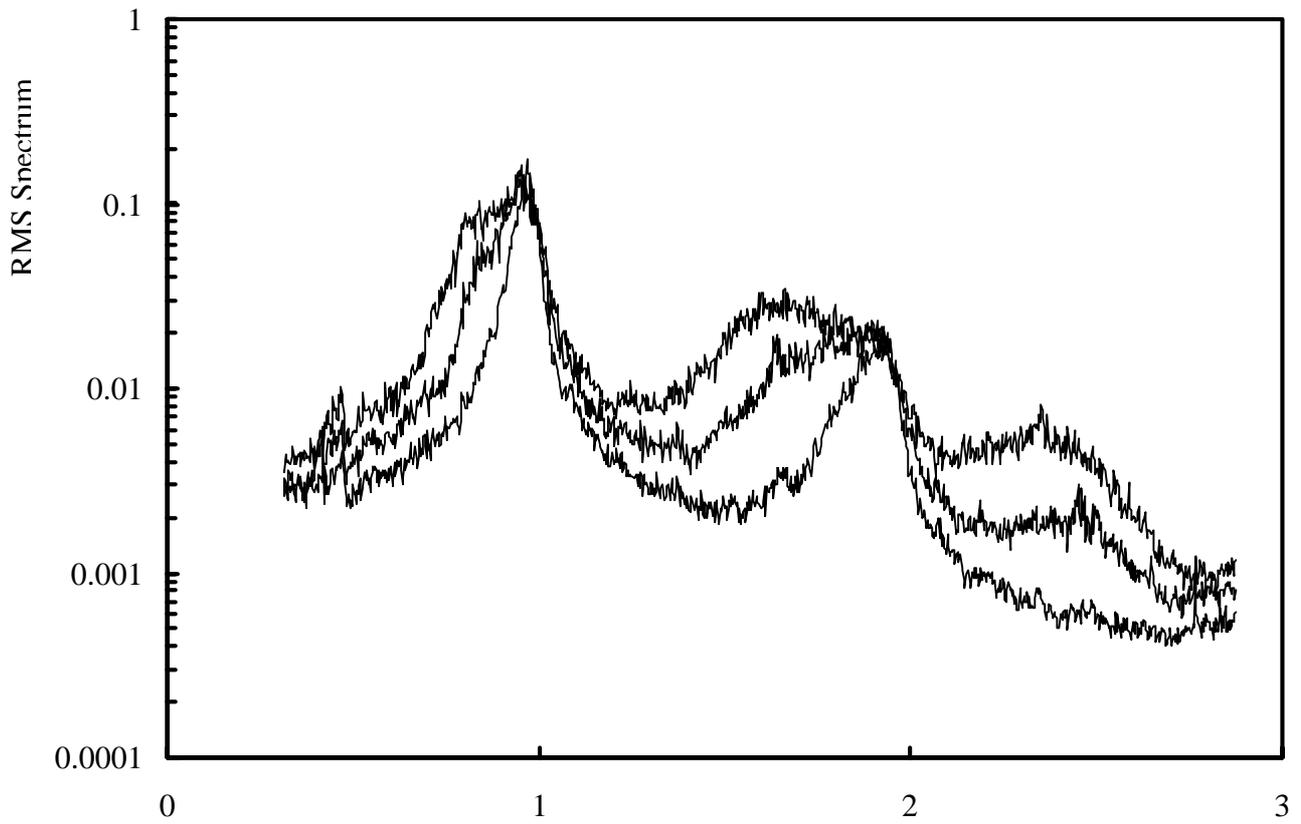

Figure 5. Experimental one sided RMS spectra of the transmitted normal force for h ≈ 6 $10^{-4}$, 1.7 $10^{-3}$ and 2.5 $10^{-3}$ (respectively a, b, c).

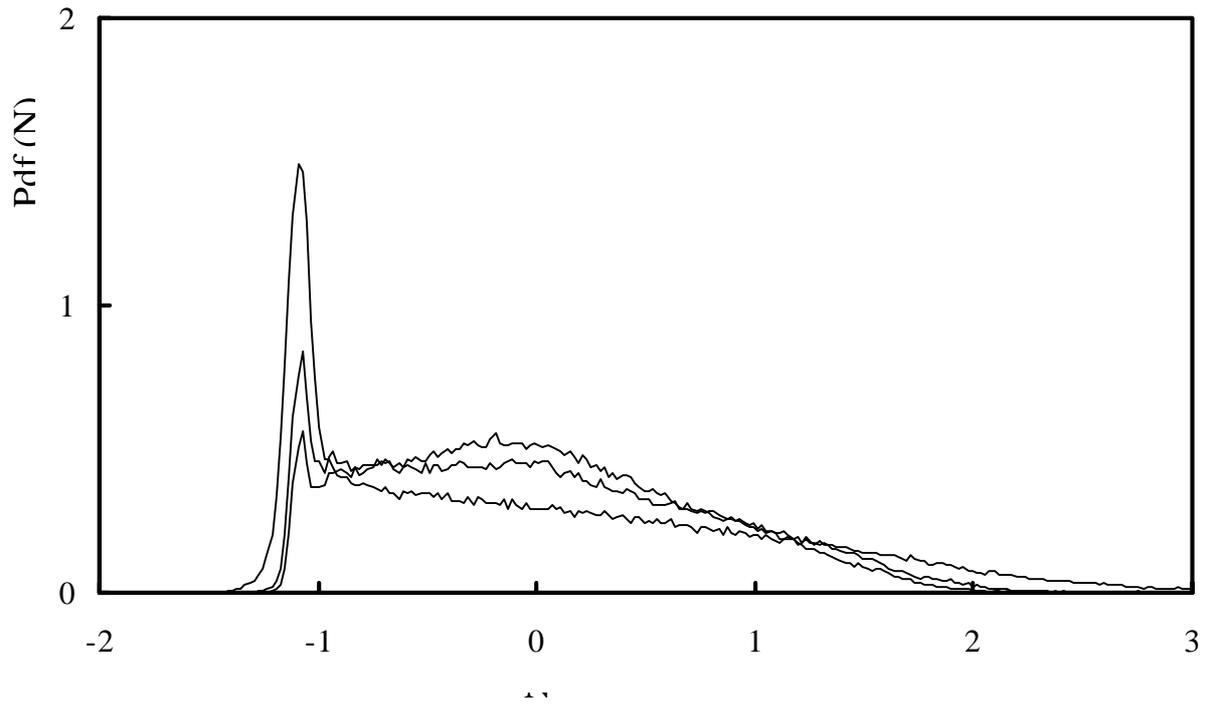

Figure 6. Probability density functions of the measured transmitted normal force for h ≈ $6 \; 10^{-4}$, $1.7 \; 10^{-3}$ and $2.5 \; 10^{-3}$ (respectively a, b, c).

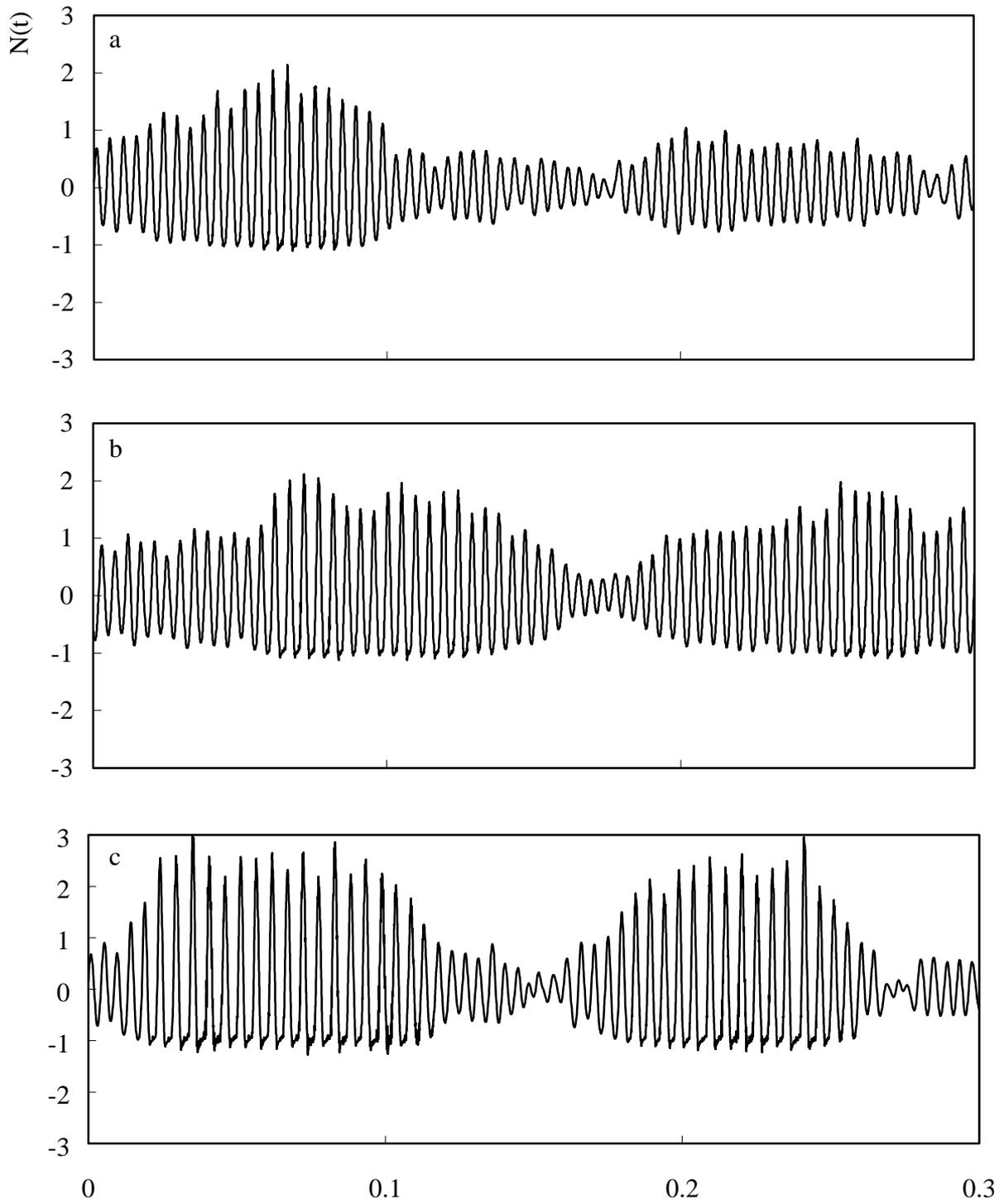

Figure 7. Time traces of the transmitted normal force for h ≈ 6 $10^{-4}$, 1.7 $10^{-3}$ and 2.5 $10^{-3}$ (respectively a, b, c).

J. PERRET-LIAUDET AND E. RIGAUD

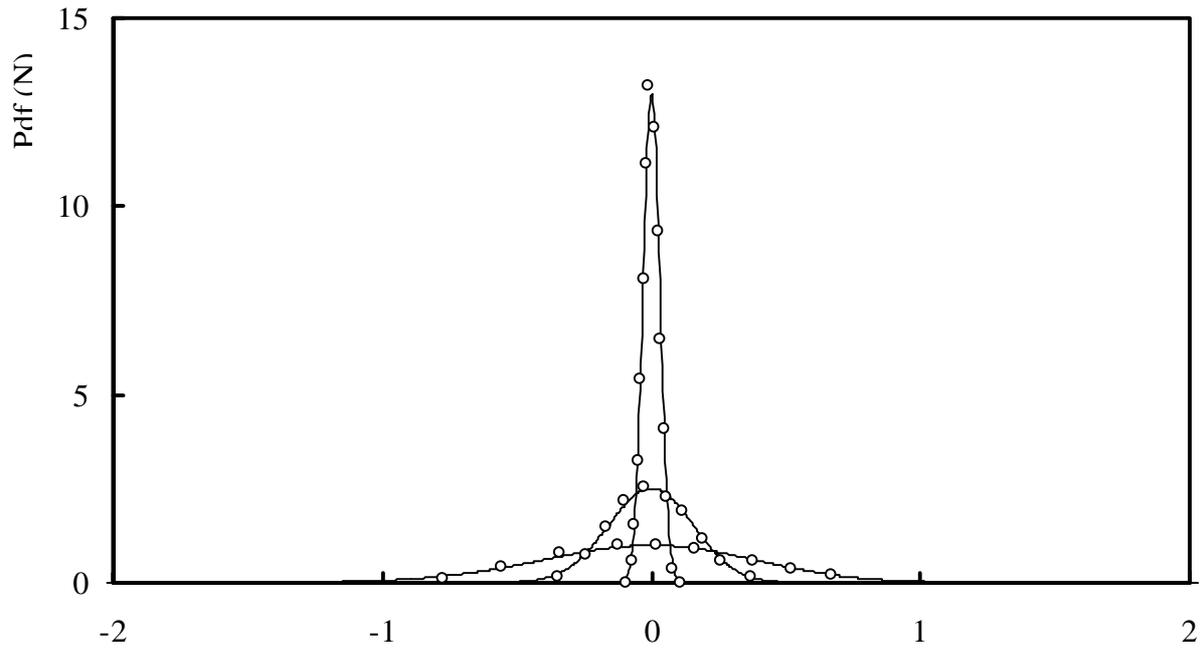

Figure 8. Probability density functions of the elastic restoring force for h = $3 \cdot 10^{-6}$, $8 \cdot 10^{-5}$ and $5 \cdot 10^{-4}$ (respectively a, b, c). Results obtained from the stationary Fokker-Planck equation ( —— ) and from Monte Carlo simulations ( ○ ).

J. PERRET-LIAUDET AND E. RIGAUD

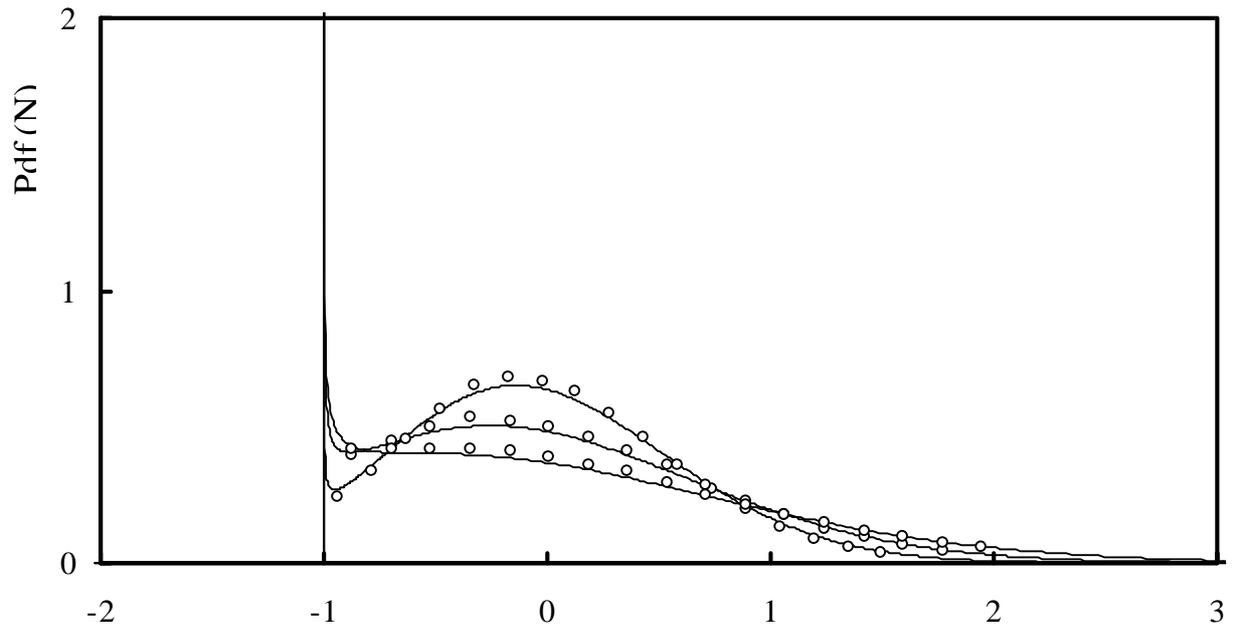

Figure 9. Probability density functions of the elastic restoring force for h = 1.2 10$^{-3}$, 2 10$^{-3}$, 3.2 10$^{-3}$ (respectively a, b, c). Results obtained from the stationary Fokker-Planck equation (——) and from Monte Carlo simulations ( O ).

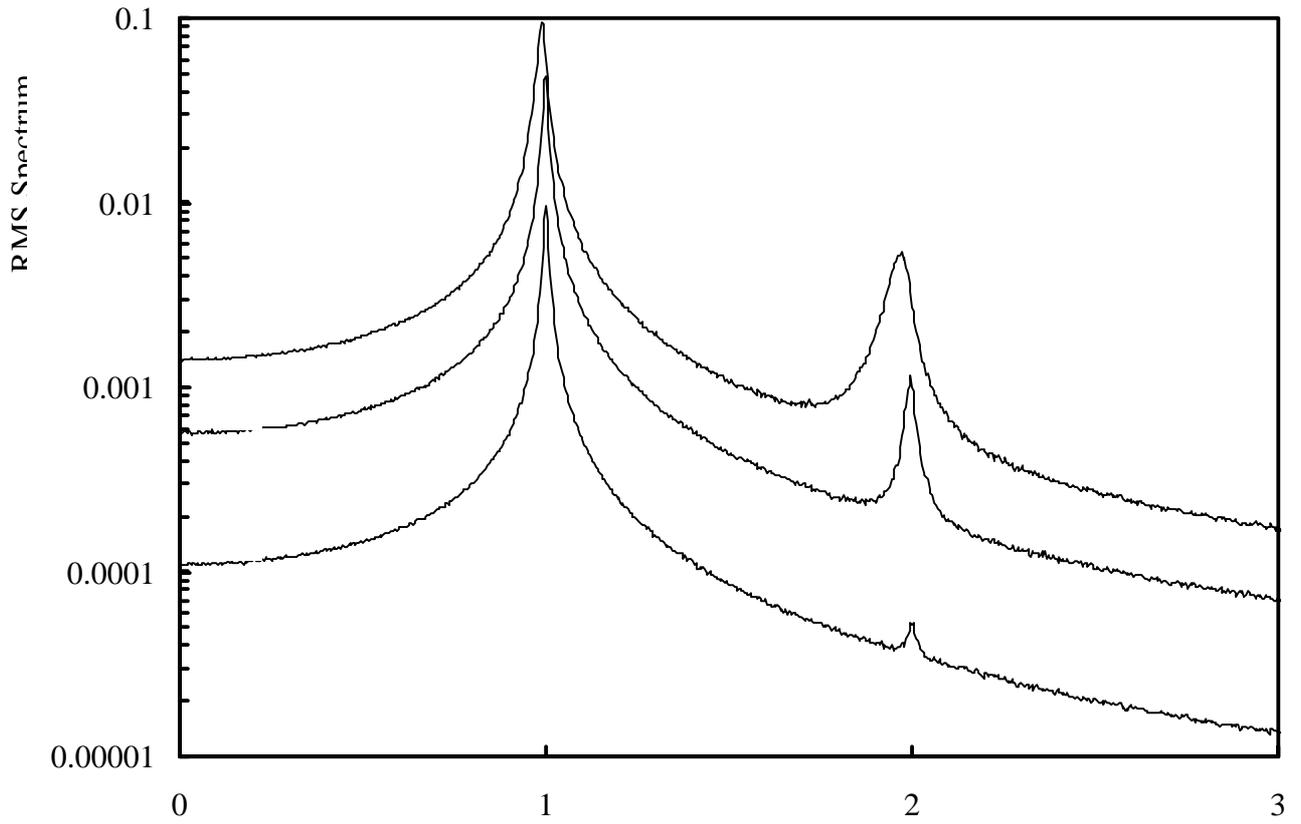

Figure 10. Numerical one sided RMS spectra of the elastic restoring force for h = $3\ 10^{-6}$, $8\ 10^{-5}$ and $5\ 10^{-4}$ (respectively a, b, c).

J. PERRET-LIAUDET AND E. RIGAUD

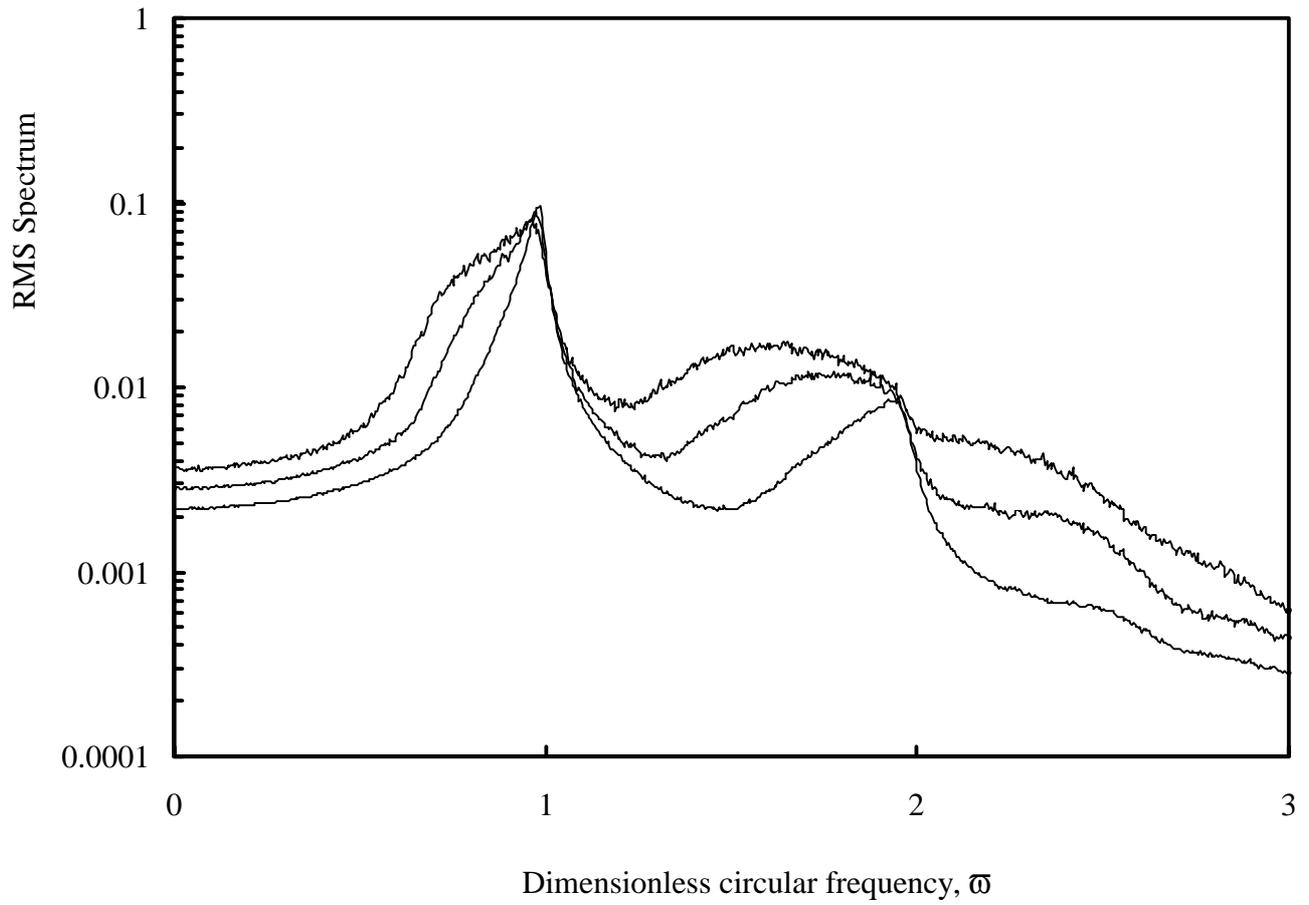

Dimensionless circular frequency, $\varpi$

Figure 11. Numerical one sided RMS spectra of the elastic restoring force for h = 1.2 $10^{-3}$, 2 $10^{-3}$ and 3.2 $10^{-3}$ (respectively a, b, c).

J. PERRET-LIAUDET AND E. RIGAUD

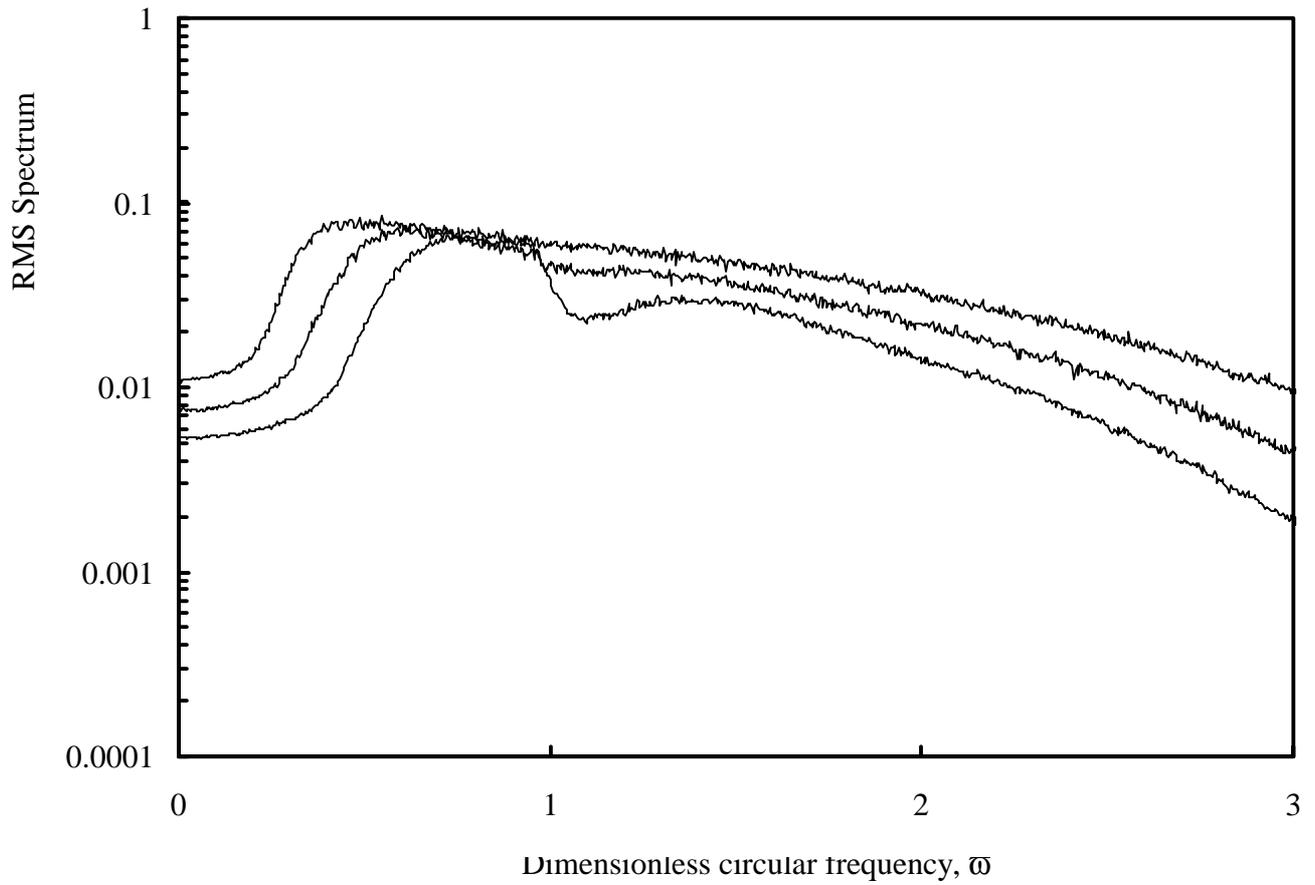

Figure 12. Numerical one sided RMS spectra of the elastic restoring force for h = 7 10$^{-3}$, 1.4 10$^{-2}$ and 3 10$^{-1}$ (respectively a, b, c).

J. PERRET-LIAUDET AND E. RIGAUD

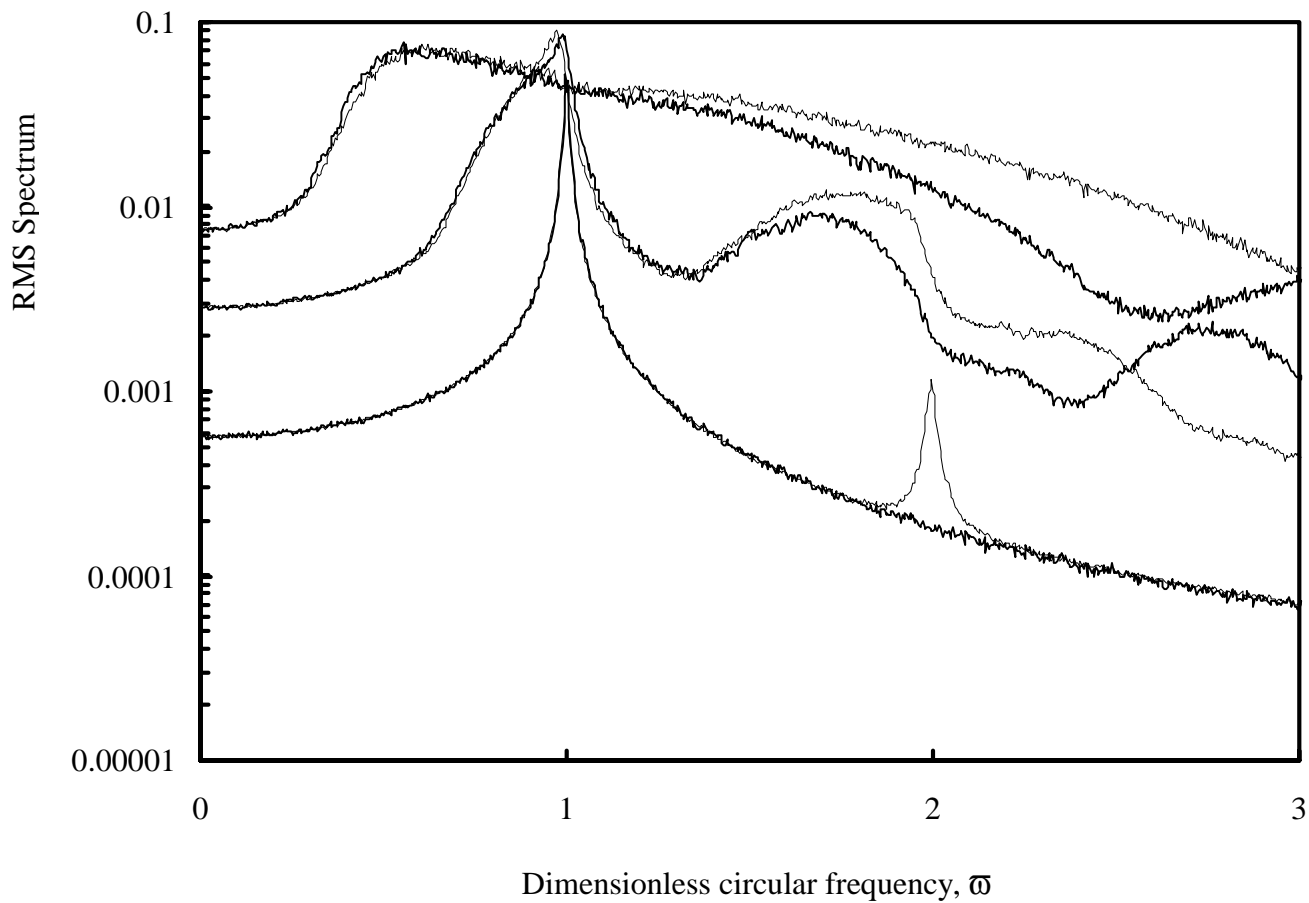

Figure 13. Numerical one sided RMS spectra of the elastic restoring force obtained with an Hertzian elastic contact law (thin line) and a linear one (thick line). $h = 8\ 10^{-5}$, $2\ 10^{-3}$ and $1.4\ 10^{-2}$ (respectively a, b, c).

J. PERRET-LIAUDET AND E. RIGAUD